\documentclass[ALICE,manyauthors]{cernphprep}
\usepackage[comma,square,numbers,sort&compress]{natbib}
\usepackage{hyperref}
\usepackage{lineno}
\usepackage{xspace}
\usepackage{lmodern}

\begin{document}

\newcommand {\red}[1]    {\textcolor{red}{#1}}
\newcommand {\green}[1]  {\textcolor{green}{#1}}
\newcommand {\blue}[1]   {\textcolor{blue}{#1}}
\newcommand {\dgreen}[1]    {\textcolor{darkgreen}{#1}}
\newcommand {\orange}[1]   {\textcolor{orange}{#1}}

\newcommand{\deutsch}[1]{\foreignlanguage{german}{#1}}

\newcommand{\CZ}[1]{\blue{{\textbf{CZ:} #1}}}
\newcommand{\PA}[1]{\red{{\textbf{PA:} #1}}}
\newcommand{\JN}[1]{\red{{\textbf{JN:} #1}}}
\newcommand{\MF}[1]{\red{{\textbf{JN:} #1}}}
\newcommand{\ToDo}      {\textcolor{blue}{\footnotesize \textsc{ToDo}}}
\newcommand{\TBC}       {\textcolor{red}{\footnotesize \textsc{TBC}}}

\DeclareRobustCommand{\unit}[2][]{%
        \begingroup%
                \def\0{#1}%
                \expandafter%
        \endgroup%
        \ifx\0\@empty%
                \ensuremath{\mathrm{#2}}%
        \else%
                \ensuremath{#1\,\mathrm{#2}}%
        \fi%
        }
\DeclareRobustCommand{\unitfrac}[3][]{%
        \begingroup%
                \def\0{#1}%
                \expandafter%
        \endgroup%
        \ifx\0\@empty%
                \raisebox{0.98ex}{\ensuremath{\mathrm{\scriptstyle#2}}}%
                \nobreak\hspace{-0.15em}\ensuremath{/}\nobreak\hspace{-0.12em}%
                \raisebox{-0.58ex}{\ensuremath{\mathrm{\scriptstyle#3}}}%
        \else
                \ensuremath{#1}\,%
                \raisebox{0.98ex}{\ensuremath{\mathrm{\scriptstyle#2}}}%
                \nobreak\hspace{-0.15em}\ensuremath{/}\nobreak\hspace{-0.12em}%
                \raisebox{-0.58ex}{\ensuremath{\mathrm{\scriptstyle#3}}}%
        \fi%
}

%
%
\newcommand{\ie}{i.\,e.\;}
\newcommand{\eg}{e.\,g.\;}

\newcommand{\run}[1]{\textsc{Run\,#1}}

%
%

\newcommand{\dNdeta}{\ensuremath{\mathrm{d}N_{\rm ch}/\mathrm{d}\eta}\xspace}

%
%
\newlength{\smallerpicsize}
\setlength{\smallerpicsize}{70mm}
\newlength{\smallpicsize}
\setlength{\smallpicsize}{90mm}
\newlength{\mediumpicsize}
\setlength{\mediumpicsize}{120mm}
\newlength{\largepicsize}
\setlength{\largepicsize}{150mm}

\newcommand{\PICX}[5]{
   \begin{figure}[!hbt]
      \begin{center}
         \vspace{3ex}
         \includegraphics[width=#3]{#1}
         \caption[#4]{\label{#2} #5}        
      \end{center}  
   \end{figure}
}

\newcommand{\PICH}[5]{
   \begin{figure}[H]
      \begin{center}
         \vspace{3ex}
         \includegraphics[width=#3]{#1}
         \caption[#4]{\label{#2} #5}        
      \end{center}  
   \end{figure}
}

%
%
%
\newcommand{\figs}{Figs.\xspace}
\newcommand{\Figs}{Figures\xspace}
\newcommand{\eqn}{equation\xspace}
\newcommand{\Eqn}{Equation\xspace}
\newcommand{\figref}[1]{Fig.~\ref{#1}}
\newcommand{\Figref}[1]{Figure~\ref{#1}}
\newcommand{\tabref}[1]{Tab.~\ref{#1}}
\newcommand{\Tabref}[1]{Table~\ref{#1}}
\newcommand{\appref}[1]{appendix~\ref{#1}}
\newcommand{\Appref}[1]{Appendix~\ref{#1}}
\newcommand{\secs}{Secs.\xspace}
\newcommand{\Secs}{Sections\xspace}
\newcommand{\secref}[1]{Sec.~\ref{#1}}
\newcommand{\Secref}[1]{Section~\ref{#1}}
\newcommand{\chaps}{Chaps.\xspace}
\newcommand{\Chaps}{Chapters\xspace}
\newcommand{\chapref}[1]{Chap.~\ref{#1}}
\newcommand{\Chapref}[1]{Chapter~\ref{#1}}
\newcommand{\lstref}[1]{Listing~\ref{#1}}
\newcommand{\Lstref}[1]{Listing~\ref{#1}}
%
%
\newcommand{\otoprule}{\midrule[\heavyrulewidth]}
\topfigrule
%
\newcommand {\stat}     {({\it stat.})~}
\newcommand {\syst}     {({\it syst.})~}
 \newcommand {\mom}       {\ensuremath{p}}
\newcommand {\pT}        {\pt}
\newcommand {\meanpT}    {\ensuremath{\langle p_{\mathrm{T}} \kern-0.1em\rangle}\xspace}
\newcommand {\mean}[1]   {\ensuremath{\langle #1 \kern-0.1em\rangle}\xspace} 
\newcommand {\sqrtsNN}   {\ensuremath{\mathbf{\sqrt{s_{\textsc{NN}}}}\xspace}}
\newcommand {\sqrts}     {\ensuremath{\mathbf{\sqrt{s}}\xspace}}
\newcommand {\vf}        {\ensuremath{v_{\mathrm{2}}}\xspace}
\newcommand {\et}        {\ensuremath{E_{\mathrm{t}}}\xspace}
\newcommand {\mT}        {\ensuremath{m_{\mathrm{T}}}\xspace}
\newcommand {\mTmZero}   {\ensuremath{m_{\mathrm{T}} - m_0}\xspace}
\newcommand {\minv}      {\mbox{$m_{\ee}$}}
\newcommand {\rap}       {\mbox{$y$}}
\newcommand {\absrap}    {\mbox{$\left | y \right | $}}
\newcommand {\rapXi}     {\mbox{$\left | y(\rmXi) \right | $}}
\newcommand {\abspseudorap} {\mbox{$\left | \eta \right | $}}
\newcommand {\pseudorap} {\mbox{$\eta$}}
\newcommand {\cTau}      {\ensuremath{c\tau}}
\newcommand {\sigee}     {$\sigma_E$/$E$}
\newcommand {\dNdy}      {\ensuremath{\mathrm{d}N/\mathrm{d}y}}
\newcommand {\dNdpt}     {\ensuremath{\mathrm{d}N/\mathrm{d}\pT }}
\newcommand {\dNdptdy}   {\ensuremath{\mathrm{d^{2}}N/\mathrm{d}\pT\mathrm{d}y }}
\newcommand {\fracdNdptdy}   {\ensuremath{ \frac{\mathrm{d^{2}}N}{\mathrm{d}\pT\mathrm{d}y } }}
\newcommand {\dNdmtdy}   {\ensuremath{\mathrm{d^{2}}N/\mathrm{d}\mT\mathrm{d}y }}
\newcommand {\dN}        {\ensuremath{\mathrm{d}N }}
\newcommand {\dNsquared} {\ensuremath{\mathrm{d^{2}}N }}
\newcommand {\dpt}       {\ensuremath{\mathrm{d}\pT }}
\newcommand {\dy}        {\ensuremath{\mathrm{d}y}}
\newcommand {\dNdyBold}  {\ensuremath{\boldsymbol{\dN/\dy}}\xspace}
\newcommand {\dNchdy}    {\ensuremath{\mathrm{d}N_\mathrm{ch}/\mathrm{d}y }\xspace}
\newcommand {\dNchdeta}  {\ensuremath{\mathrm{d}N_\mathrm{ch}/\mathrm{d}\eta }\xspace}
\newcommand {\dNchdptdeta}  {\ensuremath{\mathrm{d}N_\mathrm{ch}/\mathrm{d}\pT\mathrm{d}\eta }\xspace}
\newcommand {\Raa}       {\ensuremath{R_\mathrm{AA}}}
\newcommand {\RpPb}       {\ensuremath{R_\mathrm{pPb}}\xspace}
\newcommand {\Nevt}      {\ensuremath{N_\mathrm{evt}}}
\newcommand {\NevtINEL}  {\ensuremath{N_\mathrm{evt}(\textsc{inel})}}
\newcommand {\NevtNSD}   {\ensuremath{N_\mathrm{evt}(\textsc{nsd})}}
\newcommand{\dEdx}       {\ensuremath{\mathrm{d}E/\mathrm{d}x}\xspace}
\newcommand{\ttof}       {\ensuremath{t_\mathrm{TOF}}\xspace}
\newcommand {\ee}        {\mbox{$\mathrm {e^+e^-}$}\xspace}
\newcommand {\ep}        {\mbox{$\mathrm {e^-p}$}\xspace}
\newcommand {\pp}        {\mbox{$\mathrm {p\kern-0.05em p}$}\xspace}
\newcommand {\ppBoldMath} {\mbox{$\mathrm { \mathbf p\kern-0.05em \mathbf p }$}\xspace}
\newcommand {\ppbar}     {\mbox{$\mathrm {p\overline{p}}$}\xspace}
\newcommand {\PbPb}      {\ensuremath{\mbox{Pb--Pb}}\xspace}
\newcommand {\AuAu}      {\ensuremath{\mbox{Au--Au}}\xspace}
\newcommand {\CuCu}      {\ensuremath{\mbox{Cu--Cu}}\xspace}
\renewcommand {\AA}      {\ensuremath{\mbox{A--A}}\xspace}
\newcommand {\pA}        {\ensuremath{\mbox{p--A}}\xspace}
\newcommand {\pPb}       {\ensuremath{\mbox{p--Pb}}\xspace}
\newcommand {\Pbp}       {\ensuremath{\mbox{Pb--p}}\xspace}
\newcommand {\hPM}       {\ensuremath{h^{\pm}}\xspace}
\newcommand {\rphi}      {\ensuremath{(r,\phi)}\xspace}
\newcommand {\alphaS}    {\ensuremath{ \alpha_s}\xspace}
\newcommand {\MeanNpart} {\mbox{\ensuremath{< \kern-0.15em N_{part} \kern-0.15em >}}}

\newcommand {\sig}       {\ensuremath{S}\xspace}
\newcommand {\expsig}    {\ensuremath{\hat{S}}\xspace}
\newcommand {\prob}      {\ensuremath{P}\xspace}
\newcommand {\prior}     {\ensuremath{C}\xspace}
\newcommand {\prop}      {\ensuremath{F}\xspace}
\newcommand {\atrue}     {\ensuremath{\vec{A}_{\mathrm{true}}}\xspace}
\newcommand {\ameas}     {\ensuremath{\vec{A}_{\mathrm{meas}}}\xspace}
\newcommand {\detresp}   {\ensuremath{R}\xspace}

\newcommand {\pid}       {\ensuremath{\mathrm{\epsilon}_\mathrm{PID}}\xspace}
\newcommand {\nsigma}    {\ensuremath{\mathrm{n_{\sigma}}}\xspace}
\newcommand {\ylab} {\ensuremath{\mathrm{| y_{lab} |}}\xspace}

%
%
\newcommand {\mass}     {\mbox{\rm MeV$\kern-0.15em /\kern-0.12em c^2$}}
\newcommand {\tev}      {\mbox{${\rm TeV}$}\xspace}
\newcommand {\gev}      {\mbox{${\rm GeV}$}\xspace}
\newcommand {\mev}      {\mbox{${\rm MeV}$}\xspace}
\newcommand {\kev}      {\mbox{${\rm keV}$}\xspace}
\newcommand {\tevBoldMath}  {\mbox{${\rm \mathbf{TeV}}$}}
\newcommand {\gevBoldMath}  {\mbox{${\rm \mathbf{GeV}}$}}
\newcommand {\mmom}     {\mbox{\rm MeV$\kern-0.15em /\kern-0.12em c$}}
\newcommand {\gmom}     {\mbox{\rm GeV$\kern-0.15em /\kern-0.12em c$}}
\newcommand {\mmass}    {\mbox{\rm MeV$\kern-0.15em /\kern-0.12em c^2$}}
\newcommand {\gmass}    {\mbox{\rm GeV$\kern-0.15em /\kern-0.12em c^2$}}
\newcommand {\nb}       {\mbox{\rm nb}}
\newcommand {\musec}    {\mbox{$\mu {\rm s}$}}
\newcommand {\nsec}     {\mbox{${\rm ns}$}}
\newcommand {\psec}     {\mbox{${\rm ps}$}}
\newcommand {\fmC}      {\mbox{${\rm fm/c}$}}
\newcommand {\fm}       {\mbox{${\rm fm}$}}
\newcommand {\cm}       {\mbox{${\rm cm}$}}
\newcommand {\mm}       {\mbox{${\rm mm}$}}
\newcommand {\mim}      {\mbox{$ \mu {\rm m}$}}
\newcommand {\cmq}      {\mbox{${\rm cm}^{2}$}}
\newcommand {\mmq}      {\mbox{${\rm mm}^{2}$}}
\newcommand {\dens}     {\mbox{${\rm g}/{\rm cm}^{3}$}}
\newcommand {\lum}      {\, \mbox{${\rm cm}^{-2} {\rm s}^{-1}$}}
\newcommand {\barn}     {\, \mbox{${\rm barn}$}}
\newcommand {\m}        {\, \mbox{${\rm m}$}}
\newcommand {\dg}       {\mbox{$\kern+0.1em ^\circ$}}
\newcommand{\mpp}{\ensuremath{\mathrm{pp}}\xspace}
\newcommand{\rts}{\ensuremath{\sqrt{s}}\xspace}
\newcommand{\GeV}{\ensuremath{\mathrm{GeV}}\xspace}
\newcommand{\TeV}{\ensuremath{\mathrm{TeV}}\xspace}
\newcommand{\gevc}{\ensuremath{\mathrm{GeV}/c}\xspace}
\newcommand{\GeVc}{\gevc}
\newcommand{\mevc}{\ensuremath{\mathrm{MeV}/c}\xspace}
\newcommand{\mevcc}{\ensuremath{\mathrm{MeV}/c^{2}}\xspace}
\newcommand{\gevcc}{\ensuremath{\mathrm{GeV}/c^{2}}\xspace}
\newcommand{\pt}{\ensuremath{p_{\rm T}}\xspace}
\newcommand{\kt}{\ensuremath{k_{\rm T}}\xspace}
\newcommand {\lumi}{\mathcal{L}_{\rm int}\xspace}
\newcommand{\nbinv}{\ensuremath{\rm nb^{-1}}}
\newcommand {\ubinv}{\ensuremath{\mu\rm b^{-1}}}
\newcommand {\um}{\ensuremath{\mu\rm m}\xspace}
\newcommand{\mub}{\ensuremath{\mu\rm b}\xspace}
\newcommand{\lt}{\textless}
\newcommand{\ctau}{\ensuremath{c\tau}\xspace}

%
%

\newcommand{\ePlusMinus}       {\mbox{$\mathrm {e^{\pm}}$}\xspace}
\newcommand{\muPlusMinus}      {\mbox{$\mathrm {\mu^{\pm}}$}\xspace}

\newcommand{\pion}            {\mbox{$\mathrm {\pi}$}\xspace}
\newcommand{\piZero}            {\mbox{$\mathrm {\pi^0}$}\xspace}
\newcommand{\piMinus}           {\ensuremath{\mathrm {\pi^-}}\xspace}
\newcommand{\piPlus}            {\ensuremath{\mathrm {\pi^+}}\xspace}
\newcommand{\piPlusMinus}       {\mbox{$\mathrm {\pi^{\pm}}$}\xspace}

\newcommand{\proton}    {\mbox{$\mathrm {p}$}\xspace}
\newcommand{\pbar}      {\mbox{$\mathrm {\overline{p}}$}\xspace}
\newcommand{\pOuPbar}   {\mbox{$\mathrm {p^{\pm}}$}\xspace}
\newcommand{\DZero}     {\ensuremath{\mathrm {D^0}}\xspace}
\newcommand{\DZerobar}  {\ensuremath{\mathrm {\overline{D}^0}}\xspace}
\newcommand{\Bminus}    {\ensuremath{\mathrm {B^-}}\xspace}
\newcommand{\BZero}     {\ensuremath{\mathrm {B^0}}\xspace}
\newcommand{\BZerobar}  {\ensuremath{\mathrm {\overline{B}^0}}\xspace}

\newcommand{\Dmes}       {\ensuremath{\mathrm {D}}\xspace}

\newcommand{\Lb}{\ensuremath{\rm {\Lambda_b}}\xspace}
\newcommand{\Xic}         {\ensuremath{\mathrm {\Xi_{c}}}\xspace}
\newcommand{\lambdab}     {\ensuremath{\mathrm {\Lambda_{b}^{0}}}\xspace}
\newcommand{\lambdac}     {\ensuremath{\mathrm {\Lambda_{c}^{+}}}\xspace}
\newcommand{\xicz}        {\ensuremath{\mathrm {\Xi_{c}^{0}}}\xspace}
\newcommand{\xiczp}        {\ensuremath{\mathrm {\Xi_{c}^{0,+}}}\xspace}
\newcommand{\xicp}        {\ensuremath{\mathrm {\Xi_{c}^{+}}}\xspace}
\newcommand{\xib}        {\ensuremath{\mathrm {\Xi_{b}}}\xspace}
\newcommand{\LambdaParticle}        {\ensuremath{\mathrm {\Lambda}}\xspace}

\newcommand{\rmLambdaZ}         {\ensuremath{\mathrm {\Lambda}}\xspace}
\newcommand{\rmAlambdaZ}        {\ensuremath{\mathrm {\overline{\Lambda}}}\xspace}
\newcommand{\rmLambda}          {\ensuremath{\mathrm {\Lambda}}\xspace}
\newcommand{\rmAlambda}         {\ensuremath{\mathrm {\overline{\Lambda}}}\xspace}
\newcommand{\rmLambdas}         {\ensuremath{\mathrm {\Lambda \kern-0.2em + \kern-0.2em \overline{\Lambda}}}\xspace}

\newcommand{\Vzero}             {\ensuremath{\mathrm {V^0}}\xspace}
\newcommand{\Vzerob}             {\ensuremath{{\bold \mathrm {V^0}}}\xspace}
\newcommand{\Kzero}             {\ensuremath{\mathrm {K^0}}\xspace}
\newcommand{\Kzs}               {\ensuremath{\mathrm {K^0_S}}\xspace}
\newcommand{\phimes}            {\ensuremath{\mathrm {\phi}}\xspace}
\newcommand{\Kminus}            {\ensuremath{\mathrm {K^-}}\xspace}
\newcommand{\Kplus}             {\ensuremath{\mathrm {K^+}}\xspace}
\newcommand{\Kstar}             {\ensuremath{\mathrm {K^*}}\xspace}
\newcommand{\Kplusmin}          {\ensuremath{\mathrm {K^{\pm}}}\xspace}
\newcommand{\Jpsi}              {\ensuremath{\rm J/\psi}\xspace}
\newcommand{\DtoKpi}{\ensuremath{\rm D^0\to K^-\pi^+}\xspace}
\newcommand{\DtoKpipi}{\ensuremath{\rm D^+\to K^-\pi^+\pi^+}\xspace}
\newcommand{\DstartoDpi}{\ensuremath{\rm D^{*+}\to D^0\pi^+}\xspace}
\newcommand{\Dzero}{\ensuremath{\mathrm {D^0}}\xspace}
\newcommand{\Dzerobar}{\ensuremath{\mathrm{\overline{D}^0}}\xspace}
\newcommand{\Dstar}{\ensuremath{\rm D^{*+}}\xspace}
\newcommand{\Dplus}{\ensuremath{\rm D^+}\xspace}
\newcommand{\Ds}{\ensuremath{\rm D_s^+}\xspace}
\newcommand{\Dsubs}{\ensuremath{\rm D_{s}^+}\xspace}
\newcommand{\decleng}{\ensuremath{\rm L}_{xyz}}
\newcommand{\Lcminus}{\ensuremath{\rm {\overline{\Lambda}{}_c^-}}\xspace}
\newcommand{\Lcplus}{\lambdac}
\newcommand{\Lc}         {\Lcplus}
\newcommand{\LcD} {\ensuremath{\lambdac/\Dzero}\xspace}

\newcommand{\Lbzero}{\ensuremath{\rm {\Lambda_b^0}}\xspace}
\newcommand{\LctopKpi}{\ensuremath{\rm \Lambda_{c}^{+}\to p K^-\pi^+}\xspace}
\newcommand{\LcpKpi}{\LctopKpi}
\newcommand{\LbtoLc}{\ensuremath{\rm \Lambda_{b}^{0}\to \Lc + \rm{X}}\xspace}
\newcommand{\LctopKzS}{\ensuremath{\rm \Lambda_{c}^{+}\to p K^{0}_{S}}\xspace}
\newcommand{\LcpKs}{\LctopKzS}
\newcommand{\LctoenuLambda}{\ensuremath{\rm \Lambda_{c}^{+}\to e^{+} \nu_{e} \Lambda}\xspace}
\newcommand{\cosP}{\ensuremath{\rm cos_{\Theta_{pointing}}}\xspace}
\newcommand{\KzStopippim}{\ensuremath{\rm K^{0}_{S}\to \pi^{+} \pi^{-}}\xspace}
\newcommand{\Lambdatoppim}{\ensuremath{\rm \Lambda \to p \pi^{-}}\xspace}
\newcommand{\nue}{$\nu_e$}
\newcommand{\DtopiKzs}{\ensuremath{\rm D^+\to \pi^+ K^{0}_{S}}\xspace}
\newcommand{\DstoKKzs}{\ensuremath{\rm D_s^+\to K^+ K^{0}_{S}}\xspace}

\newcommand{\ptLc}{\ensuremath{p_{\rm T, \Lambda_c}}\xspace}
\newcommand{\ptpion}{\ensuremath{p_{\rm T, \pi}}\xspace}
\newcommand{\ptK}{\ensuremath{p_{\rm T, K}}\xspace}
\newcommand{\ptproton}{\ensuremath{p_{\rm T, \proton}}\xspace}

\newcommand{\sqrtsseven}{\ensuremath{\mathbf{\sqrt{\textit s}} = 7~\TeV}\xspace}
\newcommand{\sqrtsfive}{\ensuremath{\mathbf{\sqrt{\textit s}} = 5.02~\TeV}\xspace}
\newcommand{\sqrtsNNfive}{\ensuremath{\mathbf{\sqrt{{\textit s}_\mathrm{NN}}} = 5.02~\TeV}\xspace}

\begin{titlepage}
\PHyear{2020}       
\PHnumber{217}      
\PHdate{10 November}  

\title{\Lcplus~production and baryon-to-meson ratios in pp and \pPb collisions at \sqrtsNNfive at the LHC}
\ShortTitle{\Lcplus production in pp and \pPb collisions at \sqrtsNNfive}   

\Collaboration{ALICE Collaboration\thanks{See Appendix~\ref{app:collab} for the list of collaboration members}}
\ShortAuthor{ALICE Collaboration} 

\begin{abstract}
The prompt production of the charm baryon \Lcplus and the \LcD production 
ratios were measured at midrapidity with the \-ALICE detector 
in \pp and \pPb collisions at \sqrtsNNfive. 
These new measurements show a clear decrease of the \LcD ratio with increasing transverse momentum (\pt) in both collision systems in the range $2<\pt<12$~\GeVc, exhibiting similarities with the light-flavour baryon-to-meson ratios $\proton/\pi$ and $\Lambda / \Kzs$.
At low $\pt$, predictions that include additional colour-reconnection mechanisms beyond the leading-colour approximation; assume the existence of additional higher-mass charm-baryon states; or include hadronisation via coalescence can describe the data, while predictions driven by charm-quark fragmentation processes measured in \ee and \ep collisions significantly underestimate the data.
The results presented in this letter provide significant evidence that the established assumption of universality (colliding-system independence) of parton-to-hadron fragmentation is not sufficient to describe charm-baryon production in hadronic collisions at LHC energies.

\end{abstract}
\end{titlepage}

\setcounter{page}{2} 



Heavy-flavour hadron production in hadronic collisions occurs through the fragmentation of a charm or beauty quark, created in hard parton-parton scattering processes, into a given meson or baryon. Theoretical calculations of heavy-flavour production generally utilise the QCD factorisation theorem~\cite{Collins:1989gx}, which describes the hadron cross section as the convolution of three terms: the parton distribution functions, the parton hard-scattering cross sections, and the fragmentation functions.
It is generally assumed that the fragmentation functions are universal between collision systems and energies, and the measurement of the relative production of different heavy-flavour hadron species is sensitive to fragmentation functions used in pQCD-based calculations. While perturbative calculations at next-to-leading order with next-to-leading-log resummation~\cite{Kniehl:2005mk,Kniehl:2012ti,Cacciari:1998it,Cacciari:2012ny} generally describe the D- and B-meson cross-section measurements~\cite{Andronic:2015wma,Aaij:2015bpa,Khachatryan:2016csy,Aaij:2017qml,Acharya:2019mgn} and the ratios of strange and non-strange D mesons~\cite{Andronic:2015wma,Acharya:2019mgn} within uncertainties, heavy-flavour baryon production is less well understood.

The \Lc production cross section in pp collisions at \sqrtsseven and \pPb collisions at \sqrtsNNfive was reported by ALICE~\cite{Acharya:2017kfy}. It was shown that in both collision systems the \pT-differential \lambdac production cross section is higher than predictions from pQCD calculations with charm fragmentation tuned on previous \ee and \ep measurements~\cite{Kniehl:2005mk,Kniehl:2012ti}. The \LcD ratio in pp and \pPb collisions is consistent in both collision systems and also significantly underestimated by several Monte Carlo (MC) generators implementing different charm-quark fragmentation processes~\cite{Skands:2014pea,Christiansen:2015yqa,Bierlich:2015rha,Bellm:2015jjp}, suggesting that the fragmentation fractions of charm quarks into different hadronic states are non-universal with respect to collision system and centre-of-mass energy.
The production of charm baryons has recently been calculated within the $k_\mathrm{T}$-factorisation approach using unintegrated gluon distribution functions and the Peterson fragmentation functions~\cite{Maciula:2018iuh}, and with the GM-VFNS scheme using updated fragmentation functions from OPAL and Belle~\cite{Kniehl:2020szu}. These approaches are unable to simultaneously describe ALICE and LHCb data with the same set of parameters, suggesting that the independent parton fragmentation scheme is insufficient to fully describe the results.
An alternative explanation has been offered by a statistical hadronisation model, taking into account an augmented list of charm-baryon states based on guidance from the Relativistic Quark Model (RQM)~\cite{Ebert:2011kk} and lattice QCD~\cite{He:2019tik}, which is able to reproduce the \LcD ratio measured by ALICE.
The magnitude of the relative yields of $\Lbzero$ baryons and beauty mesons in pp collisions measured by LHCb~\cite{Aaij:2011jp,Aaij:2015fea,Aaij:2019pqz} and CMS~\cite{Chatrchyan:2012xg} offers further evidence that the fragmentation fractions in the beauty sector also vary between collision systems.

The measurement of baryon production has also been important in heavy-ion collisions, where the high energy density and temperature create a colour-deconfined state of matter~\cite{Busza:2018rrf}. A measured enhancement of the light-flavour~\cite{Abelev:2013xaa,Adams:2006wk} and charm~\cite{Acharya:2018ckj,Adam:2019hpq,Sirunyan:2019fnc} baryon-to-meson ratio at the LHC and RHIC can be explained via an additional mechanism of hadronisation known as coalescence (or recombination), where soft quarks from the medium recombine to form a meson or baryon~\cite{Fries:2008hs}, in addition to hydrodynamical radial flow. Measurements in \pPb collisions are crucial to provide an `intermediate' collision system where the generated particle multiplicities and
energy densities are between those generated in pp and A--A collisions. 
ALICE and CMS reported an enhancement of the baryon-to-meson ratios in the light-flavour sector (p/$\pi$ and $\Lambda$/$\Kzs$) at intermediate \pT ($2<\pt<10~\GeVc$) in high-multiplicity pp and \pPb collisions, similar to that observed in heavy-ion collisions~\cite{Adam:2016dau,Khachatryan:2016yru}. This adds to the evidence that small systems also exhibit collective behaviour, which may have similar physical origins in pp,
p--A, and A--A collisions~\cite{Nagle:2018nvi}. It has been suggested that hadronisation of charm quarks via coalescence may also occur in pp and \pPb collisions~\cite{Song:2018tpv,Li:2017zuj,Minissale:2020bif}.

In this letter, the measurements of the prompt production of the charm baryon \lambdac in pp collisions at \sqrtsfive in $|y|< 0.5$ and in \pPb collisions at \sqrtsNNfive in $-0.96 <\textit{y} < 0.04$ are presented, with a focus on the \LcD production ratios. The measurement is performed as an average of the \lambdac and its charge conjugate \Lcminus, collectively referred to as \lambdac in the following.
Two hadronic decay channels were measured: \LctopKpi(branching ratio BR $= 6.28\pm0.33\%$), and \LctopKzS (BR $= 1.59\pm0.08\%$)\cite{PDG20}, which were reconstructed exploiting the topology of the weakly-decaying $\Lc$ ($c\tau = 60.7 \mu$m)~\cite{PDG20}. The results from both decay channels were averaged to obtain more precise production cross sections. With respect to the results presented in~\cite{Acharya:2017kfy}, this work studies a different centre-of-mass energy for \pp collisions, and the cross section is measured in finer \pt intervals and over a wider \pt range. The overall precision of the measurements is significantly improved, by a factor of 1.5--2, depending on \pt, for both \pp and \pPb collisions. 
For a detailed description of the analysis techniques, corrections, systematic uncertainty determination, and supplementary measurements, the reader is referred to~\cite{ALICE:2020wla}.

\label{sec: Data samples and experiment}

A description of the ALICE detector and its performance are reported in~\cite{Aamodt:2008zz, Abelev:2014ffa}. The pp data sample was collected in 2017 and the \pPb data sample was collected in 2016 during the LHC Run 2.
Both pp and \pPb collisions were recorded using a Minimum Bias (MB) trigger, which required coincident signals in the two V0 scintillator detectors located on either side of the interaction vertex. 
Further offline selection was applied in order to remove background from beam--gas collisions and other machine-induced backgrounds.
To reduce superposition of more than one interaction within the colliding bunches
(pile-up), events with multiple reconstructed primary vertices were rejected. 
Only events with a $z$-coordinate of the
reconstructed vertex position within 10 cm from the nominal interaction point were used.
With these requirements, approximately one billion MB-triggered \pp events were selected,
corresponding to an integrated luminosity of $\lumi$ = 19.5 \nbinv ($\pm$ 2.1\%~\cite{ALICE-PUBLIC-2018-014}). Approximately 600 million MB-triggered \pPb events were selected, corresponding to $\lumi$ = 287 \ubinv ($\pm$ 3.7\%~\cite{Abelev:2014epa}).


The analysis techniques used for the results presented here
are described in detail in~\cite{ALICE:2020wla}. 
Charged-particle tracks and particle decay vertices are reconstructed in the central barrel using the Inner Tracking System (ITS) and the Time Projection Chamber (TPC), which are located inside a solenoid magnet of field strength 0.5\,T. In order to reduce the large combinatorial background, selections on the \Lc candidates were made based on the particle identification (PID) signals and the displacement of the decay tracks from the collision point. The PID was performed using information on the specific energy loss of charged particles as they pass through the gas of the TPC and, where available, with flight-time measurements given by the Time-Of-Flight detector (TOF).  

For the \LcpKpi analysis, candidates were built by reconstructing triplets of tracks with the correct configuration of charges. For this analysis, the high-resolution tracking provided by the detectors meant that the decay vertex of the \Lc candidates could be resolved from the interaction point.
To identify each of the p, K, and $\pi$ daughter tracks, information from the TPC and TOF was combined using the `maximum-probability'  
Bayesian approach described in~\cite{Adam:2016acv}. 
Kinematic selections were made on the \pt of the decay products of the \Lc, and geometrical selections were made on topological properties related to the displaced vertex of the \Lc decay.

The reconstruction of \LcpKs candidates relied on reconstructing the V-shaped decay of the \Kzs meson into two pions, which was then combined with a proton track (bachelor).
In \pp collisions, candidates were further selected using criteria related to PID and properties of the \LcpKs decay.
The Bayesian probability of the combined TPC and TOF response for the bachelor track to be a proton was required to be above 80\%.
The selection criteria on kinematical and geometrical variables 
included the distance of closest approach between the decay daughters, the invariant mass, and the cosine of the pointing angle of the neutral decay vertex (\Kzs) 
to the primary vertex.  

For the \LctopKzS decay channel in \pPb collisions, the analysis was performed using a multivariate technique based on the Boosted Decision Tree (BDT) algorithm provided by the Toolkit for Multivariate Data Analysis (TMVA)~\cite{Hocker:2007ht}. The BDT algorithm was trained using signal and background \LctopKzS decay candidates simulated using \mbox{PYTHIA 6.4.25}~\cite{Sjostrand:2006za} with the Perugia2011 tune~\cite{Skands:2009zm}, and the underlying \pPb event simulated with HIJING 1.36~\cite{Wang:1991hta}. 
Candidates obtained with the same reconstruction strategy previously described were preselected using loose geometrical selections and PID selection on the bachelor proton track. The model was trained independently for each \pt interval analysed, with input variables comprising the \pt and Bayesian PID probability of the proton track, the \ctau and invariant mass of the \Kzs, and the impact parameters of the \Lc decay tracks to the primary vertex.  
This model was then applied on data, and a selection on the output response was chosen based on the expected maximum significance determined from simulations.

For both decay channels the yield of \Lc baryons was extracted in each \pt interval via fits to the candidate invariant-mass distributions. The fitting function consisted of a Gaussian to estimate the signal and an exponential or polynomial function to estimate the background. The width of the Gaussian was fixed in each \pt interval to values obtained from Monte Carlo simulations, and the mean was treated as a free parameter.
A statistical significance higher than 4 standard deviations was achieved in all \pt intervals.

 
Several corrections were applied to the measurement of the \Lc cross section. 
The geometrical acceptance of the detector as well as the selection and reconstruction efficiencies for prompt \Lc were taken into account. These correction factors were determined from pp collisions generated with \mbox{PYTHIA 6} and \mbox{PYTHIA 8.243}~\cite{Sjostrand:2007gs}, with each event including either a $\rm c \bar{c}$ or a $\rm b\bar{b}$ pair. For \pPb collisions, this was supplemented with an underlying event from the HIJING event generator. In \pPb collisions the efficiency was calculated after reweighting the events based on their charged particle multiplicity. This accounts for the fact that
the event multiplicity in simulation does not reproduce the one in data, and the efficiency depends on the multiplicity of the event as a consequence of the improvement of the resolution of the primary vertex and thus of the performance of the topological selections at higher multiplicities. 
The fraction of the \Lc yield originating from beauty decays (feed-down) was obtained using the beauty-quark production cross section from FONLL~\cite{Cacciari:1998it,Cacciari:2012ny}, the fraction of beauty quarks that fragment into beauty hadrons $\mathrm{H_b}$ from LHCb measurements~\cite{Aaij:2019pqz}, and $\mathrm{H_b} \rightarrow \Lc +$ X decay kinematics from \mbox{PYTHIA 8}, as well as the selection and reconstruction efficiency of \Lc from beauty-hadron decays. The fraction of the \Lc yield from beauty decays was found to be 2\% at low $\pt$ and up to $16\%$ at high $\pt$, and was subtracted from the measured yield.
As done in the D-meson analysis~\cite{Adam:2016ich}, the possible modification of beauty-hadron production in \pPb collisions was included in the feed-down calculation by scaling the beauty-quark production by a nuclear modification factor $R_{\textrm{pPb}}^{\textrm{feed-down}}$, where it was assumed that $R_{\textrm{pPb}}^{\textrm{feed-down}}=R_{\textrm{pPb}}^{\textrm{prompt}}$ with their ratio varied in the range $ 0.9 < R_{\textrm{pPb}}^{\textrm{feed-down}}/R_{\textrm{pPb}}^{\textrm{prompt}} < 1.3$ to evaluate the systematic uncertainties.

Systematic uncertainties on the \Lc cross sections were estimated considering the same sources as described in~\cite{Acharya:2017kfy}.
The contributions from the raw-yield extraction were evaluated by repeating the fits varying the fit interval and the functional form of the background fit function. For each of these variations the four combinations of free and fixed Gaussian mean and width parameters of the fit were considered. Overall, the relative uncertainty ranged from 4\% to 11\% depending on the \pt and analysis.
The uncertainties on the track reconstruction efficiency were estimated by adding in quadrature the uncertainty due to track quality selection and the uncertainty due to the TPC-ITS matching efficiency (from 3\% to 7\%).
The former is estimated by varying the track-quality selection criteria
and the latter is estimated by comparing the probability to match the tracks from the TPC to the ITS hits in data and simulation.
The uncertainty on the \Lc selection efficiency was estimated by varying the selection on the kinematical and topological properties of the \Lc decays, or the selection on the BDT response (from 3\% to 15\%).
The uncertainty on the PID efficiency was estimated by varying the selection on the Bayesian probability variables (from 2\% to 5\%). 
The systematic effect on the efficiencies due to the shape of the simulated \Lc \pt distribution was evaluated by reweighting the generated \Lc from \mbox{PYTHIA 6} to match the \pt distribution obtained from FONLL calculations for D mesons (maximum 1\% uncertainty).
The relative statistical uncertainty on the acceptance and efficiency correction was considered as an additional systematic uncertainty source (from 1--2\% at low \pt to 3--5\% at high \pt).
The uncertainties on $f_{\rm prompt}$ were estimated by varying the hypothesis on the production of \Lc from B-hadron decays to account for the theoretical uncertainties of b-quark production within FONLL and experimental uncertainties on B-hadron fragmentation (around 2\% at low \pt, and from 4\% to 7\% at high \pt, depending on the analysis). 
Global uncertainties of the measurement include those from the luminosity and $\Lc$ branching ratios.
The raw-yield extraction uncertainty source are considered to be uncorrelated across $\pt$ bins, while all other sources are considered to be correlated.

The results in each collision system from the two \Lc decay channels were averaged to obtain the final results. A weighted average of the results was calculated, with weights defined as the inverse of the quadratic sum of the relative statistical and uncorrelated systematic uncertainties.
The sources of systematic uncertainty assumed to be uncorrelated between different decay channels were those due to the raw-yield extraction, the statistical uncertainties on the efficiency and acceptance, and those related to the \Lc selection. The remaining uncertainties were assumed to be correlated, except the branching ratio uncertainties, which were treated as partially correlated among the hadronic-decay modes as defined in~\cite{PDG20}.

\begin{figure*}[ht!]
	\centering
	\includegraphics[width=0.49\textwidth]{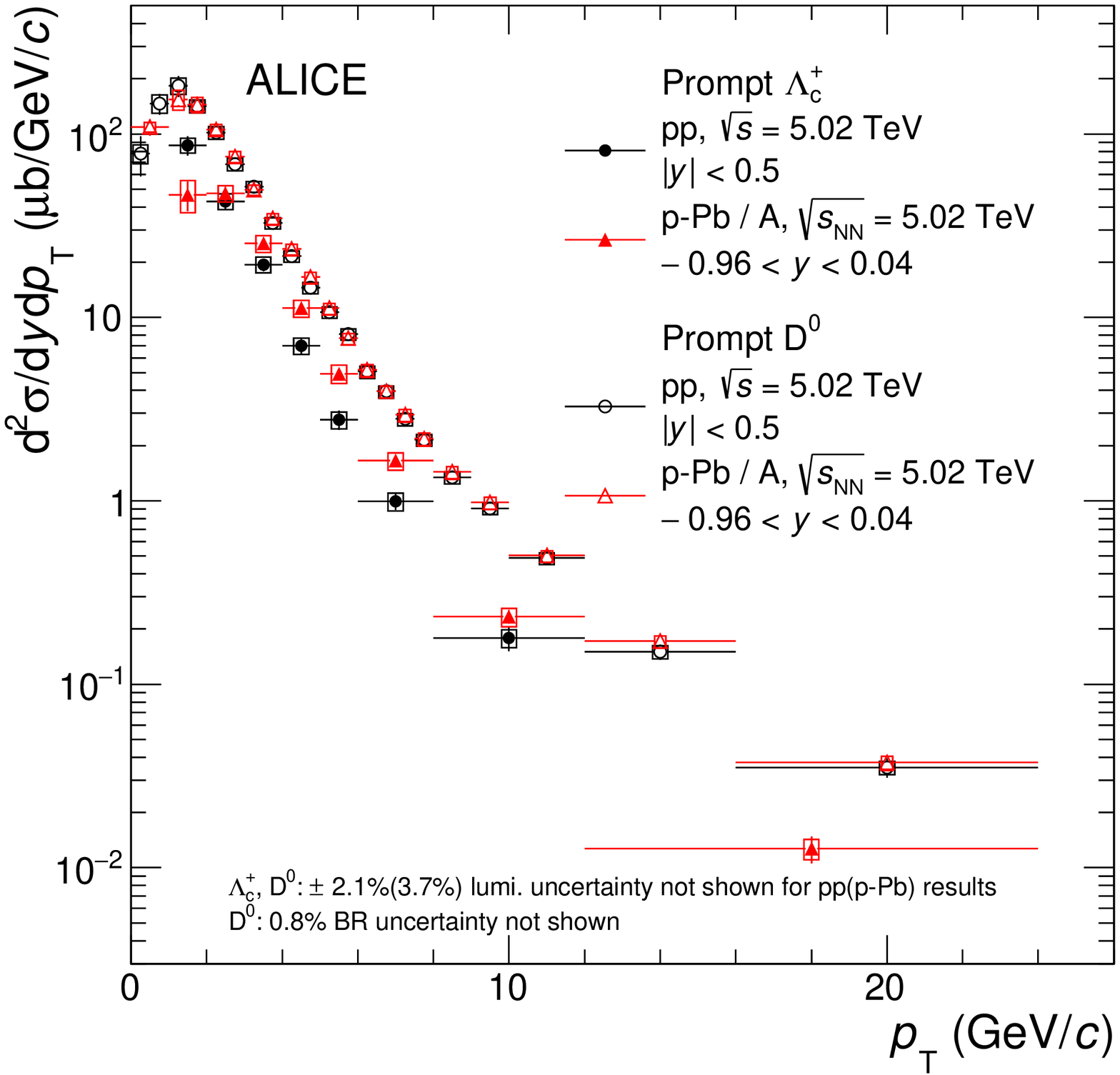}
	\includegraphics[width=0.49\textwidth]{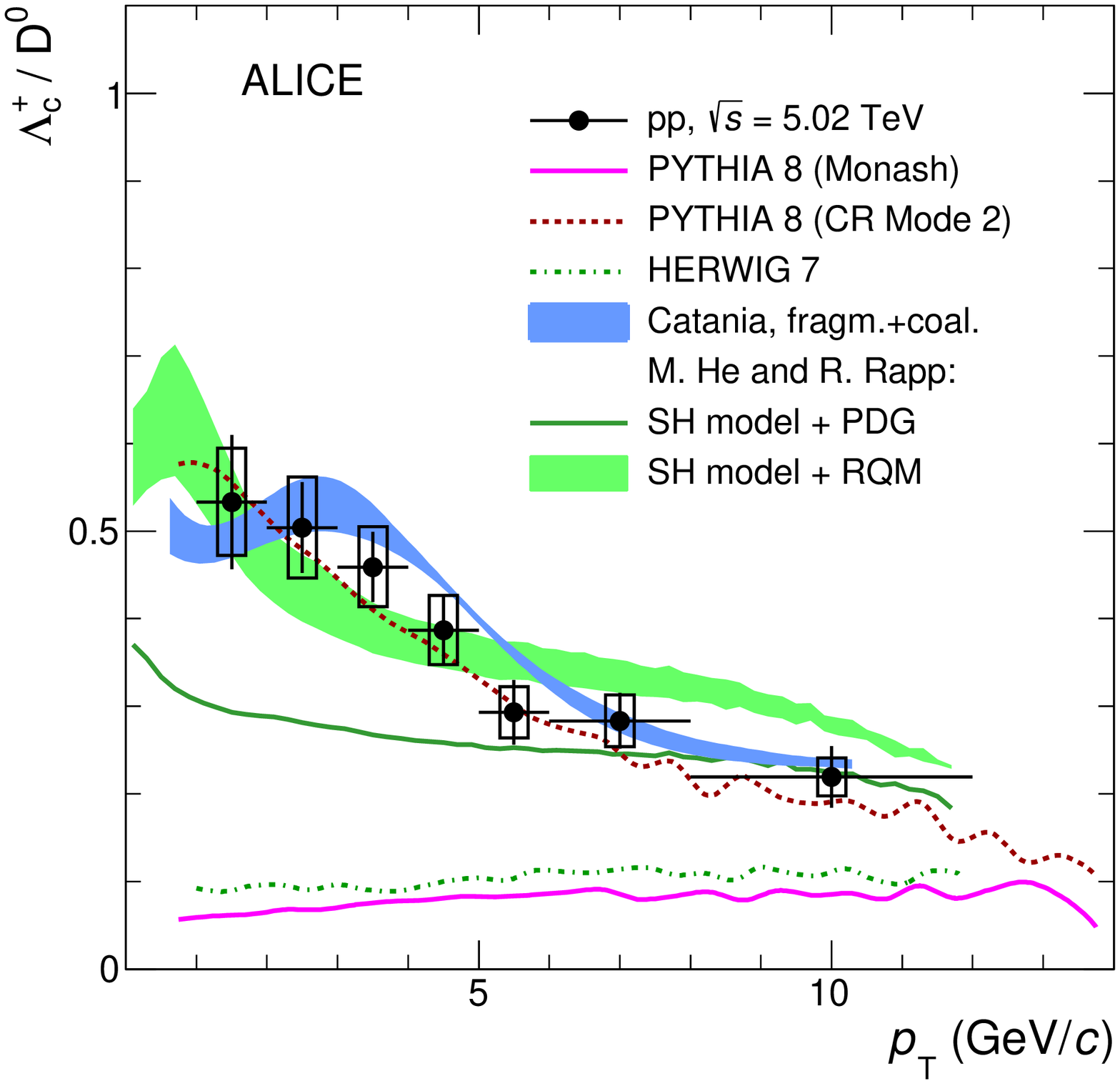}
	\caption{Left: Prompt \Lcplus and \Dzero \pt-differential cross section in pp collisions and in \pPb collisions at~\sqrtsNNfive. The results in \pPb collisions are scaled with the atomic mass number $A$ of the Pb nucleus. Right: the \LcD ratio as a function of \pt measured in pp collisions at \sqrtsfive compared with theoretical predictions (see text for details). Statistical uncertainties are shown as vertical bars, while systematic uncertainties are shown as boxes, and the bin widths are shown as horizontal bars.}		
	\label{fig:LcD-withModels}
\end{figure*}

\Figref{fig:LcD-withModels} (left) shows a comparison of the \Lcplus \pt-differential cross sections in pp and in \pPb collisions at \sqrtsNNfive. The \Dzero \pt-differential cross sections measured in the same collision systems and at the same centre-of-mass energy during the same data taking periods~\cite{Acharya:2019mgn,Acharya:2019mno} are also shown. In order to compare the spectral shapes in the two different collision systems at the same energy, the results in \pPb collisions are scaled by the atomic mass number of the lead nucleus. 
For \Lc baryons the spectral shape in \pPb collisions is slightly harder than in \pp collisions, while for $\Dzero$ mesons the spectral shapes are fully consistent within uncertainties.

\Figref{fig:LcD-withModels} (right) shows the baryon-to-meson ratio \LcD measured in pp collisions at \sqrtsfive
as a function of \pt, compared to theoretical predictions. The uncertainty on the luminosity cancels in the ratio. The $\LcD$ ratio is measured to be 0.4--0.5 at low \pt, and decreases to around 0.2 at high \pt. 
The previous results at \sqrtsseven hinted at a decrease of the \LcD ratio with \pt,  
although the precision was not enough to confirm this~\cite{Acharya:2017kfy}. The results in \pp collisions at \sqrtsfive, with much higher precision than \sqrtsseven results, show a clear decrease with increasing \pt.
The strong \pt-dependence of the \LcD ratio is in contrast with the ratios of strange and non-strange D mesons in pp collisions at \sqrtsfive and \sqrtsseven~\cite{Acharya:2019mgn,Acharya:2017jgo} and in \pPb collisions at \sqrtsNNfive~\cite{Acharya:2019mno}, which do not show a significant \pt dependence within uncertainties and thus indicate that there are no large differences between fragmentation functions of charm quarks to charm mesons.
The result presented here instead provides strong indications that the fragmentation functions of baryons and mesons differ significantly. 

The measured \LcD ratios in \pp collisions are compared with predictions from several MC generators and models in which different hadronisation processes are implemented. The \mbox{PYTHIA 8} predictions include the Monash tune~\cite{Skands:2014pea} and a tune that implements colour reconnection beyond the leading-colour approximation, corresponding to CR Mode 2 as defined in~\cite{Christiansen:2015yqa}.  
Hadronisation in PYTHIA is built on the Lund string fragmentation model~\cite{ANDERSSON198331,Andersson:1998tv}, where quarks and gluons connected by colour strings fragment into hadrons, and colour reconnection allows for partons created in the collision to interact via colour strings. The latter tune introduces new colour reconnection topologies beyond the leading-colour approximation, including `junctions' that fragment into baryons, leading to increased baryon production.
As a technical point, the PYTHIA~8 simulations are generated with all soft QCD processes switched on~\cite{Sjostrand:2007gs}. 
The PYTHIA~8 Monash tune and HERWIG 7.2~\cite{Bellm:2015jjp} predictions are driven by the fragmentation fraction $f(\mathrm{c} \rightarrow \Lc)$ implemented in these generators, which all suggest a relatively constant \LcD ratio versus \pt of about 0.1, significantly underestimating the data at low \pT. At high \pt, the data approach the predictions from these generators, although the measurement in $8 < \pt < 12$ \gevc is still underestimated by about a factor of 2.
A significant enhancement of the \LcD ratio is seen with colour reconnection beyond the leading-colour approximation (PYTHIA 8 CR Mode 2).
This prediction is consistent with the measured \LcD ratio in pp collisions, also reproducing the downward \pt trend.
The statistical hadronisation model (`SH model' in the legend)~\cite{He:2019tik} uses either an underlying charm-baryon spectrum taken from the PDG, or includes additional excited charm baryons that have not yet been observed but are predicted by the RQM. These additional states decay strongly to \Lc baryons, which contribute to the prompt \Lc spectrum. 
The RQM predictions include a source of uncertainty related to the branching ratios of the excited baryon states into \Lc final states, which is estimated by varying the branching ratios between 50\% and 100\%. With the PDG charm-baryon spectrum 
the model underpredicts the data. With the additional baryon states 
the model instead gives a good description of the pp data, both in the magnitude of the ratio, and the decreasing trend with \pt. 
The Catania model~\cite{Minissale:2020bif} assumes that a colour-deconfined state of matter is formed and hadronisation can occur via coalescence in addition to fragmentation. Coalescence is implemented through the Wigner formalism, where a blast wave model is used to determine the $\pt$ spectrum of light quarks and FONLL pQCD calculations are used for heavy quarks. Hadronisation via coalescence is predicted to dominate at low $\pt$, while fragmentation dominates at high $\pt$. This model provides a good description of both the magnitude and shape of the data over the full $\pt$ range.

\begin{figure*}[ht!]
	\centering
	\includegraphics[width=0.95\textwidth]{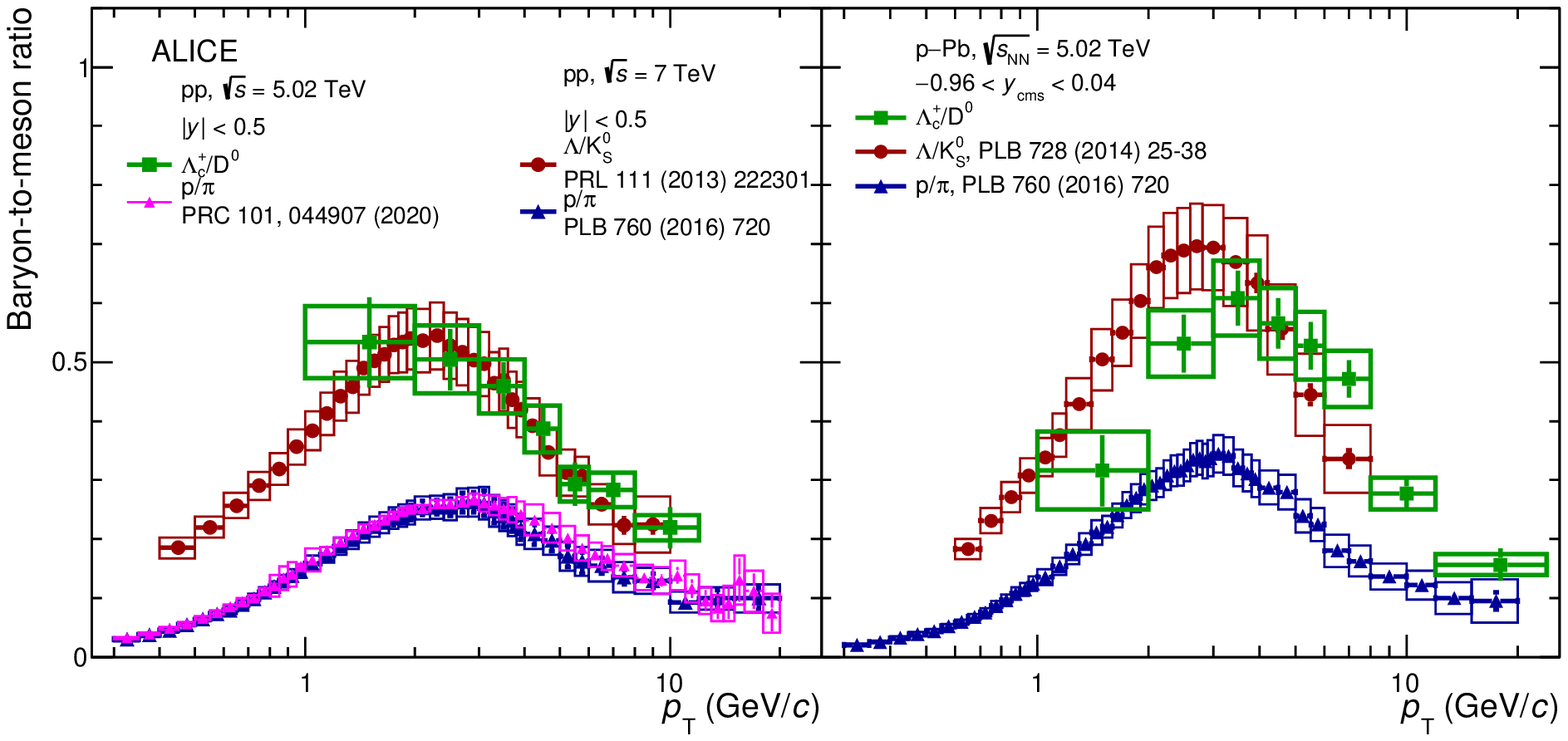} 
	\caption{The charm baryon-to-meson ratio \LcD in pp collisions (left), and \pPb collisions (right) at \sqrtsNNfive, compared to the light-flavour baryon-to-meson ratios $\Lambda / \Kzs$ and $\proton/\pi$. Statistical uncertainties are shown as vertical bars, while systematic uncertainties are shown as boxes, and the bin widths are shown as horizontal bars.}
	\label{fig:BaryonMesonLF}
\end{figure*}

\Figref{fig:BaryonMesonLF} shows the \LcD baryon-to-meson ratio measured in pp collisions at \sqrtsfive (left) and in \pPb collisions at \sqrtsNNfive (right) as a function of \pt, compared to baryon-to-meson ratios in the light-flavour sector, $\Lambda / \Kzs$~\cite{Abelev:2013xaa,Abelev:2013haa} and $\proton / \pion$~\cite{Adam:2016dau,Acharya:2019yoi} (calculated as the sum of both charged particles and antiparticles, $(\mathrm{p} + \bar{\mathrm{p}})/(\pi^+ + \pi^-)$). The p/$\pi$ ratio
in pp collisions is shown at both \sqrtsfive and \sqrtsseven, displaying consistent results at both centre-of-mass energies, while the $\Lambda/\Kzs$ ratio in pp collisions is shown only at \sqrtsseven. Unlike heavy-flavour hadron production, which occurs primarily through the fragmentation of a charm quark produced in the initial hard scattering, light-flavour hadrons have a significant contribution from gluon fragmentation. Low-\pt light-flavour hadrons also primarily originate from soft scattering processes involving small momentum transfers. All particle yields in these ratios were corrected for feed-down from weak decays, although the pion spectrum is expected to have significant feed-down contributions also from the strong decays of other particle species, primarily $\rho$ and $\omega$ mesons.
Despite these differences, the three ratios, \LcD, $\Lambda / \Kzs$, and $\proton / \pion$ demonstrate some remarkably similar characteristics in both collision systems. All ratios exhibit a decreasing trend after $\pt \gtrsim2$--3~\gevc. The \LcD and $\Lambda / \Kzs$ ratios are consistent, in terms of both shape and magnitude, within uncertainties. The light-flavour ratios both peak at ${\sim}$2--3 \gevc in both \pp and \pPb collisions, and there is an indication of a peak at $2 < \pt < 4~\gevc$ in the \LcD ratio in \pPb collisions. These similarities between heavy-flavour and light-flavour measurements hint at a potential common mechanism for light- and charm-baryon formation in \pp and \pPb collisions at LHC energies.
It is interesting to note that all baryon-to-meson ratios also indicate a shift toward higher momenta in \pPb collisions, which for light-flavour particle production is often attributed to radial flow~\cite{Abelev:2013haa}. However, while flow effects in the charm sector (\Dzero and heavy-flavour decay leptons) have been observed in high-multiplicity \pPb collisions~\cite{Sirunyan:2018toe,Acharya:2018dxy}, these effects are expected to be smaller at lower multiplicities, and also smaller for charm than for light-flavour hadrons.

In summary, \Lc-baryon production was measured in \pp collisions at midrapidity ($|y| < 0.5$) and in \pPb collisions in the rapidity interval $-0.96 < y < 0.04$ at \sqrtsNNfive. 
A clear \pT-dependence of the \Lc/\Dzero ratio is reported, with the ratio decreasing as the \pT increases.
This trend is similar to that of baryon-to-meson ratios measured in the light-flavour sector in \pp and \pPb collisions, suggesting common mechanisms for light- and charm-baryon formation.
While models incorporating fragmentation parameters from \ee and \ep collisions significantly underestimate the 
\LcD ratio, three models can reproduce the measurements. The first is a tune of PYTHIA~8 which considers that, in \pp collisions at high energy, multi-parton interactions produce a rich hadronic environment that requires an extension of colour reconnection in hadronisation processes beyond the leading-colour approximation. The second method is the SH+RQM model, which relies on the presence of a large set of yet-unobserved higher-mass charm-baryon states with relative yields following the Statistical Hadronisation model. The third relies on hadronisation via coalescence and fragmentation after the formation of a colour-deconfined state of matter. All three models imply a substantially different description of the charm-baryon production in pp collisions with respect to \ee and \ep collisions, indicating that the assumption of universal parton-to-hadron fragmentation between collision systems is not sufficient to describe charm-baryon production.


\newenvironment{acknowledgement}{\relax}{\relax}
\begin{acknowledgement}
\section*{Acknowledgements}

The ALICE Collaboration would like to thank all its engineers and technicians for their invaluable contributions to the construction of the experiment and the CERN accelerator teams for the outstanding performance of the LHC complex.
The ALICE Collaboration gratefully acknowledges the resources and support provided by all Grid centres and the Worldwide LHC Computing Grid (WLCG) collaboration.
The ALICE Collaboration acknowledges the following funding agencies for their support in building and running the ALICE detector:
A. I. Alikhanyan National Science Laboratory (Yerevan Physics Institute) Foundation (ANSL), State Committee of Science and World Federation of Scientists (WFS), Armenia;
Austrian Academy of Sciences, Austrian Science Fund (FWF): [M 2467-N36] and Nationalstiftung f\"{u}r Forschung, Technologie und Entwicklung, Austria;
Ministry of Communications and High Technologies, National Nuclear Research Center, Azerbaijan;
Conselho Nacional de Desenvolvimento Cient\'{\i}fico e Tecnol\'{o}gico (CNPq), Financiadora de Estudos e Projetos (Finep), Funda\c{c}\~{a}o de Amparo \`{a} Pesquisa do Estado de S\~{a}o Paulo (FAPESP) and Universidade Federal do Rio Grande do Sul (UFRGS), Brazil;
Ministry of Education of China (MOEC) , Ministry of Science \& Technology of China (MSTC) and National Natural Science Foundation of China (NSFC), China;
Ministry of Science and Education and Croatian Science Foundation, Croatia;
Centro de Aplicaciones Tecnol\'{o}gicas y Desarrollo Nuclear (CEADEN), Cubaenerg\'{\i}a, Cuba;
Ministry of Education, Youth and Sports of the Czech Republic, Czech Republic;
The Danish Council for Independent Research | Natural Sciences, the VILLUM FONDEN and Danish National Research Foundation (DNRF), Denmark;
Helsinki Institute of Physics (HIP), Finland;
Commissariat \`{a} l'Energie Atomique (CEA) and Institut National de Physique Nucl\'{e}aire et de Physique des Particules (IN2P3) and Centre National de la Recherche Scientifique (CNRS), France;
Bundesministerium f\"{u}r Bildung und Forschung (BMBF) and GSI Helmholtzzentrum f\"{u}r Schwerionenforschung GmbH, Germany;
General Secretariat for Research and Technology, Ministry of Education, Research and Religions, Greece;
National Research, Development and Innovation Office, Hungary;
Department of Atomic Energy Government of India (DAE), Department of Science and Technology, Government of India (DST), University Grants Commission, Government of India (UGC) and Council of Scientific and Industrial Research (CSIR), India;
Indonesian Institute of Science, Indonesia;
Istituto Nazionale di Fisica Nucleare (INFN), Italy;
Institute for Innovative Science and Technology , Nagasaki Institute of Applied Science (IIST), Japanese Ministry of Education, Culture, Sports, Science and Technology (MEXT) and Japan Society for the Promotion of Science (JSPS) KAKENHI, Japan;
Consejo Nacional de Ciencia (CONACYT) y Tecnolog\'{i}a, through Fondo de Cooperaci\'{o}n Internacional en Ciencia y Tecnolog\'{i}a (FONCICYT) and Direcci\'{o}n General de Asuntos del Personal Academico (DGAPA), Mexico;
Nederlandse Organisatie voor Wetenschappelijk Onderzoek (NWO), Netherlands;
The Research Council of Norway, Norway;
Commission on Science and Technology for Sustainable Development in the South (COMSATS), Pakistan;
Pontificia Universidad Cat\'{o}lica del Per\'{u}, Peru;
Ministry of Science and Higher Education, National Science Centre and WUT ID-UB, Poland;
Korea Institute of Science and Technology Information and National Research Foundation of Korea (NRF), Republic of Korea;
Ministry of Education and Scientific Research, Institute of Atomic Physics and Ministry of Research and Innovation and Institute of Atomic Physics, Romania;
Joint Institute for Nuclear Research (JINR), Ministry of Education and Science of the Russian Federation, National Research Centre Kurchatov Institute, Russian Science Foundation and Russian Foundation for Basic Research, Russia;
Ministry of Education, Science, Research and Sport of the Slovak Republic, Slovakia;
National Research Foundation of South Africa, South Africa;
Swedish Research Council (VR) and Knut \& Alice Wallenberg Foundation (KAW), Sweden;
European Organization for Nuclear Research, Switzerland;
Suranaree University of Technology (SUT), National Science and Technology Development Agency (NSDTA) and Office of the Higher Education Commission under NRU project of Thailand, Thailand;
Turkish Atomic Energy Agency (TAEK), Turkey;
National Academy of  Sciences of Ukraine, Ukraine;
Science and Technology Facilities Council (STFC), United Kingdom;
National Science Foundation of the United States of America (NSF) and United States Department of Energy, Office of Nuclear Physics (DOE NP), United States of America.
\end{acknowledgement}

\bibliographystyle{utphys}   
\bibliography{include/bibfile}

\newpage
\appendix

%
%

\section{The ALICE Collaboration}
\label{app:collab}
\begingroup
\small
\begin{flushleft}


S.~Acharya$^{\rm 142}$, 
D.~Adamov\'{a}$^{\rm 97}$, 
A.~Adler$^{\rm 75}$, 
J.~Adolfsson$^{\rm 82}$, 
G.~Aglieri Rinella$^{\rm 35}$, 
M.~Agnello$^{\rm 31}$, 
N.~Agrawal$^{\rm 55}$, 
Z.~Ahammed$^{\rm 142}$, 
S.~Ahmad$^{\rm 16}$, 
S.U.~Ahn$^{\rm 77}$
% ...

\section*{Affiliation notes}

$^{\rm I}$ Deceased\\
$^{\rm II}$ Also at: Italian National Agency for New Technologies, Energy and Sustainable Economic Development (ENEA), Bologna, Italy\\
$^{\rm III}$ Also at: Dipartimento DET del Politecnico di Torino, Turin, Italy\\
$^{\rm IV}$ Also at: M.V. Lomonosov Moscow State University, D.V. Skobeltsyn Institute of Nuclear, Physics, Moscow, Russia\\
$^{\rm V}$ Also at: Institute of Theoretical Physics, University of Wroclaw, Poland\\

\section*{Collaboration Institutes}

$^{1}$ A.I. Alikhanyan National Science Laboratory (Yerevan Physics Institute) Foundation, Yerevan, Armenia\\
$^{2}$ AGH University of Science and Technology, Cracow, Poland\\
$^{3}$ Bogolyubov Institute for Theoretical Physics, National Academy of Sciences of Ukraine, Kiev, Ukraine\\
$^{4}$ Bose Institute, Department of Physics  and Centre for Astroparticle Physics and Space Science (CAPSS), Kolkata, India\\
$^{5}$ Budker Institute for Nuclear Physics, Novosibirsk, Russia\\
$^{6}$ California Polytechnic State University, San Luis Obispo, California, United States\\
$^{7}$ Central China Normal University, Wuhan, China\\
$^{8}$ Centro de Aplicaciones Tecnol\'{o}gicas y Desarrollo Nuclear (CEADEN), Havana, Cuba\\
$^{9}$ Centro de Investigaci\'{o}n y de Estudios Avanzados (CINVESTAV), Mexico City and M\'{e}rida, Mexico\\
$^{10}$ Chicago State University, Chicago, Illinois, United States\\
$^{11}$ China Institute of Atomic Energy, Beijing, China\\
$^{12}$ Chungbuk National University, Cheongju, Republic of Korea\\
$^{13}$ Comenius University Bratislava, Faculty of Mathematics, Physics and Informatics, Bratislava, Slovakia\\
$^{14}$ COMSATS University Islamabad, Islamabad, Pakistan\\
$^{15}$ Creighton University, Omaha, Nebraska, United States\\
$^{16}$ Department of Physics, Aligarh Muslim University, Aligarh, India\\
$^{17}$ Department of Physics, Pusan National University, Pusan, Republic of Korea\\
$^{18}$ Department of Physics, Sejong University, Seoul, Republic of Korea\\
$^{19}$ Department of Physics, University of California, Berkeley, California, United States\\
$^{20}$ Department of Physics, University of Oslo, Oslo, Norway\\
$^{21}$ Department of Physics and Technology, University of Bergen, Bergen, Norway\\
$^{22}$ Dipartimento di Fisica dell'Universit\`{a} 'La Sapienza' and Sezione INFN, Rome, Italy\\
$^{23}$ Dipartimento di Fisica dell'Universit\`{a} and Sezione INFN, Cagliari, Italy\\
$^{24}$ Dipartimento di Fisica dell'Universit\`{a} and Sezione INFN, Trieste, Italy\\
$^{25}$ Dipartimento di Fisica dell'Universit\`{a} and Sezione INFN, Turin, Italy\\
$^{26}$ Dipartimento di Fisica e Astronomia dell'Universit\`{a} and Sezione INFN, Bologna, Italy\\
$^{27}$ Dipartimento di Fisica e Astronomia dell'Universit\`{a} and Sezione INFN, Catania, Italy\\
$^{28}$ Dipartimento di Fisica e Astronomia dell'Universit\`{a} and Sezione INFN, Padova, Italy\\
$^{29}$ Dipartimento di Fisica e Nucleare e Teorica, Universit\`{a} di Pavia  and Sezione INFN, Pavia, Italy\\
$^{30}$ Dipartimento di Fisica `E.R.~Caianiello' dell'Universit\`{a} and Gruppo Collegato INFN, Salerno, Italy\\
$^{31}$ Dipartimento DISAT del Politecnico and Sezione INFN, Turin, Italy\\
$^{32}$ Dipartimento di Scienze e Innovazione Tecnologica dell'Universit\`{a} del Piemonte Orientale and INFN Sezione di Torino, Alessandria, Italy\\
$^{33}$ Dipartimento di Scienze MIFT, Universit\`{a} di Messina, Messina, Italy\\
$^{34}$ Dipartimento Interateneo di Fisica `M.~Merlin' and Sezione INFN, Bari, Italy\\
$^{35}$ European Organization for Nuclear Research (CERN), Geneva, Switzerland\\
$^{36}$ Faculty of Electrical Engineering, Mechanical Engineering and Naval Architecture, University of Split, Split, Croatia\\
$^{37}$ Faculty of Engineering and Science, Western Norway University of Applied Sciences, Bergen, Norway\\
$^{38}$ Faculty of Nuclear Sciences and Physical Engineering, Czech Technical University in Prague, Prague, Czech Republic\\
$^{39}$ Faculty of Science, P.J.~\v{S}af\'{a}rik University, Ko\v{s}ice, Slovakia\\
$^{40}$ Frankfurt Institute for Advanced Studies, Johann Wolfgang Goethe-Universit\"{a}t Frankfurt, Frankfurt, Germany\\
$^{41}$ Fudan University, Shanghai, China\\
$^{42}$ Gangneung-Wonju National University, Gangneung, Republic of Korea\\
$^{43}$ Gauhati University, Department of Physics, Guwahati, India\\
$^{44}$ Helmholtz-Institut f\"{u}r Strahlen- und Kernphysik, Rheinische Friedrich-Wilhelms-Universit\"{a}t Bonn, Bonn, Germany\\
$^{45}$ Helsinki Institute of Physics (HIP), Helsinki, Finland\\
$^{46}$ High Energy Physics Group,  Universidad Aut\'{o}noma de Puebla, Puebla, Mexico\\
$^{47}$ Hiroshima University, Hiroshima, Japan\\
$^{48}$ Hochschule Worms, Zentrum  f\"{u}r Technologietransfer und Telekommunikation (ZTT), Worms, Germany\\
$^{49}$ Horia Hulubei National Institute of Physics and Nuclear Engineering, Bucharest, Romania\\
$^{50}$ Indian Institute of Technology Bombay (IIT), Mumbai, India\\
$^{51}$ Indian Institute of Technology Indore, Indore, India\\
$^{52}$ Indonesian Institute of Sciences, Jakarta, Indonesia\\
$^{53}$ INFN, Laboratori Nazionali di Frascati, Frascati, Italy\\
$^{54}$ INFN, Sezione di Bari, Bari, Italy\\
$^{55}$ INFN, Sezione di Bologna, Bologna, Italy\\
$^{56}$ INFN, Sezione di Cagliari, Cagliari, Italy\\
$^{57}$ INFN, Sezione di Catania, Catania, Italy\\
$^{58}$ INFN, Sezione di Padova, Padova, Italy\\
$^{59}$ INFN, Sezione di Roma, Rome, Italy\\
$^{60}$ INFN, Sezione di Torino, Turin, Italy\\
$^{61}$ INFN, Sezione di Trieste, Trieste, Italy\\
$^{62}$ Inha University, Incheon, Republic of Korea\\
$^{63}$ Institute for Gravitational and Subatomic Physics (GRASP), Utrecht University/Nikhef, Utrecht, Netherlands\\
$^{64}$ Institute for Nuclear Research, Academy of Sciences, Moscow, Russia\\
$^{65}$ Institute of Experimental Physics, Slovak Academy of Sciences, Ko\v{s}ice, Slovakia\\
$^{66}$ Institute of Physics, Homi Bhabha National Institute, Bhubaneswar, India\\
$^{67}$ Institute of Physics of the Czech Academy of Sciences, Prague, Czech Republic\\
$^{68}$ Institute of Space Science (ISS), Bucharest, Romania\\
$^{69}$ Institut f\"{u}r Kernphysik, Johann Wolfgang Goethe-Universit\"{a}t Frankfurt, Frankfurt, Germany\\
$^{70}$ Instituto de Ciencias Nucleares, Universidad Nacional Aut\'{o}noma de M\'{e}xico, Mexico City, Mexico\\
$^{71}$ Instituto de F\'{i}sica, Universidade Federal do Rio Grande do Sul (UFRGS), Porto Alegre, Brazil\\
$^{72}$ Instituto de F\'{\i}sica, Universidad Nacional Aut\'{o}noma de M\'{e}xico, Mexico City, Mexico\\
$^{73}$ iThemba LABS, National Research Foundation, Somerset West, South Africa\\
$^{74}$ Jeonbuk National University, Jeonju, Republic of Korea\\
$^{75}$ Johann-Wolfgang-Goethe Universit\"{a}t Frankfurt Institut f\"{u}r Informatik, Fachbereich Informatik und Mathematik, Frankfurt, Germany\\
$^{76}$ Joint Institute for Nuclear Research (JINR), Dubna, Russia\\
$^{77}$ Korea Institute of Science and Technology Information, Daejeon, Republic of Korea\\
$^{78}$ KTO Karatay University, Konya, Turkey\\
$^{79}$ Laboratoire de Physique des 2 Infinis, Ir\`{e}ne Joliot-Curie, Orsay, France\\
$^{80}$ Laboratoire de Physique Subatomique et de Cosmologie, Universit\'{e} Grenoble-Alpes, CNRS-IN2P3, Grenoble, France\\
$^{81}$ Lawrence Berkeley National Laboratory, Berkeley, California, United States\\
$^{82}$ Lund University Department of Physics, Division of Particle Physics, Lund, Sweden\\
$^{83}$ Moscow Institute for Physics and Technology, Moscow, Russia\\
$^{84}$ Nagasaki Institute of Applied Science, Nagasaki, Japan\\
$^{85}$ Nara Women{'}s University (NWU), Nara, Japan\\
$^{86}$ National and Kapodistrian University of Athens, School of Science, Department of Physics , Athens, Greece\\
$^{87}$ National Centre for Nuclear Research, Warsaw, Poland\\
$^{88}$ National Institute of Science Education and Research, Homi Bhabha National Institute, Jatni, India\\
$^{89}$ National Nuclear Research Center, Baku, Azerbaijan\\
$^{90}$ National Research Centre Kurchatov Institute, Moscow, Russia\\
$^{91}$ Niels Bohr Institute, University of Copenhagen, Copenhagen, Denmark\\
$^{92}$ Nikhef, National institute for subatomic physics, Amsterdam, Netherlands\\
$^{93}$ NRC Kurchatov Institute IHEP, Protvino, Russia\\
$^{94}$ NRC <<Kurchatov>> Institute - ITEP, Moscow, Russia\\
$^{95}$ NRNU Moscow Engineering Physics Institute, Moscow, Russia\\
$^{96}$ Nuclear Physics Group, STFC Daresbury Laboratory, Daresbury, United Kingdom\\
$^{97}$ Nuclear Physics Institute of the Czech Academy of Sciences, \v{R}e\v{z} u Prahy, Czech Republic\\
$^{98}$ Oak Ridge National Laboratory, Oak Ridge, Tennessee, United States\\
$^{99}$ Ohio State University, Columbus, Ohio, United States\\
$^{100}$ Petersburg Nuclear Physics Institute, Gatchina, Russia\\
$^{101}$ Physics department, Faculty of science, University of Zagreb, Zagreb, Croatia\\
$^{102}$ Physics Department, Panjab University, Chandigarh, India\\
$^{103}$ Physics Department, University of Jammu, Jammu, India\\
$^{104}$ Physics Department, University of Rajasthan, Jaipur, India\\
$^{105}$ Physikalisches Institut, Eberhard-Karls-Universit\"{a}t T\"{u}bingen, T\"{u}bingen, Germany\\
$^{106}$ Physikalisches Institut, Ruprecht-Karls-Universit\"{a}t Heidelberg, Heidelberg, Germany\\
$^{107}$ Physik Department, Technische Universit\"{a}t M\"{u}nchen, Munich, Germany\\
$^{108}$ Politecnico di Bari and Sezione INFN, Bari, Italy\\
$^{109}$ Research Division and ExtreMe Matter Institute EMMI, GSI Helmholtzzentrum f\"ur Schwerionenforschung GmbH, Darmstadt, Germany\\
$^{110}$ Rudjer Bo\v{s}kovi\'{c} Institute, Zagreb, Croatia\\
$^{111}$ Russian Federal Nuclear Center (VNIIEF), Sarov, Russia\\
$^{112}$ Saha Institute of Nuclear Physics, Homi Bhabha National Institute, Kolkata, India\\
$^{113}$ School of Physics and Astronomy, University of Birmingham, Birmingham, United Kingdom\\
$^{114}$ Secci\'{o}n F\'{\i}sica, Departamento de Ciencias, Pontificia Universidad Cat\'{o}lica del Per\'{u}, Lima, Peru\\
$^{115}$ St. Petersburg State University, St. Petersburg, Russia\\
$^{116}$ Stefan Meyer Institut f\"{u}r Subatomare Physik (SMI), Vienna, Austria\\
$^{117}$ SUBATECH, IMT Atlantique, Universit\'{e} de Nantes, CNRS-IN2P3, Nantes, France\\
$^{118}$ Suranaree University of Technology, Nakhon Ratchasima, Thailand\\
$^{119}$ Technical University of Ko\v{s}ice, Ko\v{s}ice, Slovakia\\
$^{120}$ The Henryk Niewodniczanski Institute of Nuclear Physics, Polish Academy of Sciences, Cracow, Poland\\
$^{121}$ The University of Texas at Austin, Austin, Texas, United States\\
$^{122}$ Universidad Aut\'{o}noma de Sinaloa, Culiac\'{a}n, Mexico\\
$^{123}$ Universidade de S\~{a}o Paulo (USP), S\~{a}o Paulo, Brazil\\
$^{124}$ Universidade Estadual de Campinas (UNICAMP), Campinas, Brazil\\
$^{125}$ Universidade Federal do ABC, Santo Andre, Brazil\\
$^{126}$ University of Cape Town, Cape Town, South Africa\\
$^{127}$ University of Houston, Houston, Texas, United States\\
$^{128}$ University of Jyv\"{a}skyl\"{a}, Jyv\"{a}skyl\"{a}, Finland\\
$^{129}$ University of Liverpool, Liverpool, United Kingdom\\
$^{130}$ University of Science and Technology of China, Hefei, China\\
$^{131}$ University of South-Eastern Norway, Tonsberg, Norway\\
$^{132}$ University of Tennessee, Knoxville, Tennessee, United States\\
$^{133}$ University of the Witwatersrand, Johannesburg, South Africa\\
$^{134}$ University of Tokyo, Tokyo, Japan\\
$^{135}$ University of Tsukuba, Tsukuba, Japan\\
$^{136}$ Universit\'{e} Clermont Auvergne, CNRS/IN2P3, LPC, Clermont-Ferrand, France\\
$^{137}$ Universit\'{e} de Lyon, CNRS/IN2P3, Institut de Physique des 2 Infinis de Lyon , Lyon, France\\
$^{138}$ Universit\'{e} de Strasbourg, CNRS, IPHC UMR 7178, F-67000 Strasbourg, France, Strasbourg, France\\
$^{139}$ Universit\'{e} Paris-Saclay Centre d'Etudes de Saclay (CEA), IRFU, D\'{e}partment de Physique Nucl\'{e}aire (DPhN), Saclay, France\\
$^{140}$ Universit\`{a} degli Studi di Foggia, Foggia, Italy\\
$^{141}$ Universit\`{a} di Brescia and Sezione INFN, Brescia, Italy\\
$^{142}$ Variable Energy Cyclotron Centre, Homi Bhabha National Institute, Kolkata, India\\
$^{143}$ Warsaw University of Technology, Warsaw, Poland\\
$^{144}$ Wayne State University, Detroit, Michigan, United States\\
$^{145}$ Westf\"{a}lische Wilhelms-Universit\"{a}t M\"{u}nster, Institut f\"{u}r Kernphysik, M\"{u}nster, Germany\\
$^{146}$ Wigner Research Centre for Physics, Budapest, Hungary\\
$^{147}$ Yale University, New Haven, Connecticut, United States\\
$^{148}$ Yonsei University, Seoul, Republic of Korea\\

\endgroup  
\end{document}